\documentclass[runningheads]{llncs}
\usepackage[utf8]{inputenc}
\usepackage[T1]{fontenc}
\usepackage[english]{babel}
\usepackage{amssymb,amsmath}

\usepackage[font=small,center]{caption}
\usepackage{tikz}
\usetikzlibrary{positioning,arrows,hobby,backgrounds,calc,trees}

\usepackage{graphicx}
\usepackage{rotating}
\usepackage{wrapfig}
\usepackage{makecell}
\usepackage{multirow}

\usepackage[bookmarks,unicode,colorlinks=true]{hyperref}

\makeatletter
   \def\@citecolor{blue}%
   \def\@urlcolor{blue}%
   \def\@linkcolor{blue}%

\def\orcidID#1{\smash{\href{http://orcid.org/#1}{\protect\raisebox{-1.25pt}{\protect\includegraphics{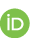}}}}}
\makeatother

\begin{document}
\title{Revisiting Semantics of Interactions for Trace Validity Analysis} 

\titlerunning{Revisiting Semantics of Interactions for Trace Validity Analysis}
\author{Erwan Mahe\inst{1}\orcidID{0000-0002-5322-4337} \and
Christophe Gaston\inst{2}\orcidID{0000-0001-6865-5108} \and
Pascale Le Gall\inst{1}\orcidID{0000-0002-8955-6835} }

\authorrunning{E. Mahe, C. Gaston, P. Le Gall}
\institute{Laboratoire de Mathématiques et Informatique pour la Complexité et les Systèmes\\
CentraleSupélec - Plateau de Moulon\\
9 rue Joliot-Curie, F-91192 Gif-sur-Yvette Cedex\and
CEA, LIST, Laboratory of Systems Requirements and Conformity Engineering, P.C. 174, Gif-sur-Yvette, 91191, France}
\maketitle
\begin{abstract}
Interaction languages such as MSC are often associated with formal semantics by means of translations into distinct behavioral formalisms such as automatas or Petri nets.
In contrast to translational approaches we propose an operational approach. Its principle is to identify which elementary communication actions can be immediately executed, and then to compute, for every such action, a new interaction representing the possible continuations to its execution. 
We also define an algorithm for checking the validity of execution traces (i.e. whether or not they belong to an interaction's semantics). Algorithms for semantic computation and trace validity are analyzed by means of experiments.
\end{abstract}

\section{Introduction}

\begin{wrapfigure}{r}{0.25\textwidth}
\vspace{-2cm}
\centering
\begin{tabular}{c}
\includegraphics[scale=0.2]{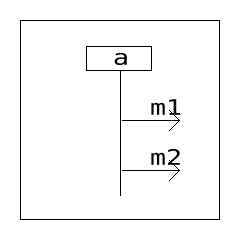}\\
{\scriptsize (a) Default sequencing}\\
{\scriptsize $i=seq(a!m1,a!m2)$}\\
\includegraphics[scale=0.2]{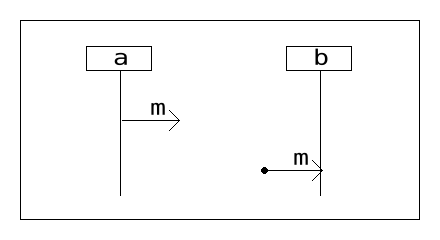}\\
{\scriptsize (b) Uncorrelated instants}\\
{\scriptsize $i=seq(a!m,b?m)$}\\
\includegraphics[scale=0.2]{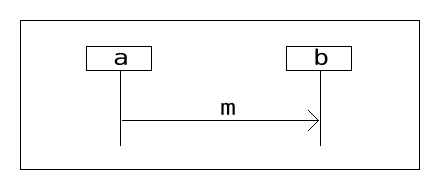}\\
{\scriptsize (c) Message passing}\\
{\scriptsize $i=strict(a!m,b?m)$}
\end{tabular}
\caption{UML-SD style}
\label{fig:umlsd_style}
\vspace{-1cm}
\end{wrapfigure}

\begin{wrapfigure}{l}{0.46\textwidth}
\vspace{-0.675cm}
\centering
\begin{tabular}{cc}
\makecell{\includegraphics[scale=0.2]{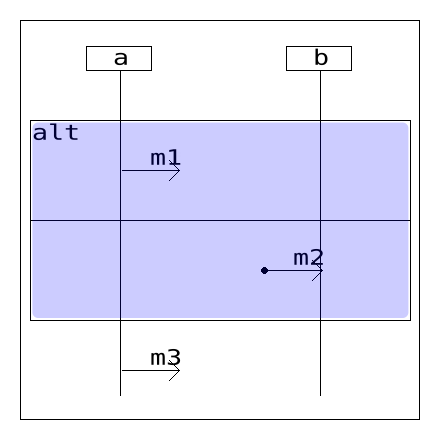}} & \makecell{\resizebox{0.225\textwidth}{!}{
            \begin{tikzpicture}[every node/.style = {shape=rectangle, align=center}]
	            \node (o) { $seq$ } [sibling distance=1.25cm,level distance=0.75cm]
            	    child {	node (o1) {	$alt$ } [sibling distance=1.25cm]
	            	    child { node (o11) { $a!m_{1}$ }
            			    }
            		    child { node (o12) { $b?m_{2}$ }
		            	    }
	            	}
	                child { node (o2) { $a!m_3$ } };
	       \begin{pgfonlayer}{background}
\draw[blue,fill=blue,opacity=0.2](o1.north east) to[closed,curve through={(o1.east) .. (o1.south east) .. (o12.north) .. (o12.north east) .. (o12.east) .. (o12.south east) .. (o12.south) .. (o12.south west) .. (o11.south east) .. (o11.south) .. (o11.south west) .. (o11.west) .. (o11.north west) .. (o11.north) .. (o1.west) .. (o1.north west)}](o1.north);
\end{pgfonlayer};
            \end{tikzpicture}
        }}\\
\multicolumn{2}{c}{
\makecell{{\scriptsize whole interaction $i = i_{|\epsilon}$}\\
{\scriptsize subinteraction $i_{|1}$ in blue}}
}
\end{tabular}
\caption{Syntax and Positions}
\label{fig:syntax_and_positions}
\vspace{-0.675cm}
\end{wrapfigure}
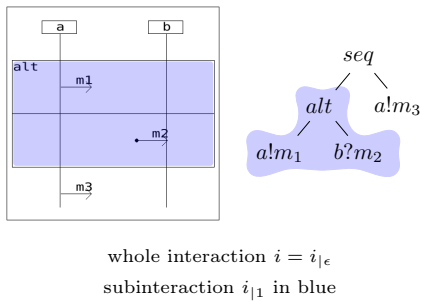

Interaction Languages (IL) are powerful mechanisms to express behavioral requirements in the form of scenarios called {\em interactions}. ILs include several recognized standards such as MSC and LSC \cite{lscs_breathing_life_into_message_sequence_charts}, HMSC \cite{MauwR97}, MSD \cite{assert_and_negate_revisited_modal_semantics_for_UML_sequence_diagrams}, UML-Sequence Diagrams \cite{the_many_meanings_of_uml2_sd_a_survey} (UML-SD), etc. These graphical languages represent parts involved in a communication scheme as vertical lines, called lifelines. Each one highlights a succession of instants where actions (emissions or receptions of messages) may occur. These instants are conventionally ordered from top to bottom as illustrated (in the style of UML-SD) in Fig.\ref{fig:umlsd_style}-a, where the emission of $m_1$ occurs before that of $m_2$. However, this sequencing does not order actions occurring on different lifelines; in Fig.\ref{fig:umlsd_style}-b, even though the reception of $m$ occurs graphically below the emission of $m$, no order is enforced. As such, this specificity is called 'weak sequencing'. In order to enforce a causality relation between such uncorrelated actions, we use a different 'strict sequencing' operator. In Fig.\ref{fig:umlsd_style}-c, it is used to express a message $m$ passing between lifelines $a$ and $b$. Here, $m$ cannot be received before being emitted; the origin of the arrow denoting an instant preceding the one depicted by its target. Additional operators (e.g. UML-SD combined fragments) enable the expression of various concepts to order actions such as parallelisation, repetition, alternatives (illustrated in Fig.\ref{fig:syntax_and_positions}), etc. They structure interactions and specify relative scheduling for subscenarii.

When ILs are fitted with formal semantics, requirements can be processed using formal techniques, such as model-checking \cite{AlurY99} or model-based testing \cite{global_and_local_testing_from_message_sequence_charts}. As pointed out earlier, the key semantic concept here is the causality relation between actions that the interaction's structure induce. Valid traces are those respecting the subsequent partial order \cite{semantics_of_interactions_in_uml_2_0,global_and_local_testing_from_message_sequence_charts}. The authors of \cite{UML_interactions_meet_state_machines_an_institutional_approach} define a simple IL as a set of terms built above basic actions and provide it with a denotational semantics which associates each interaction term with a set of traces. This kind of formal framework can serve as a reference for stating theorems about interactions (e.g. the 'satisfaction condition' proven in \cite{UML_interactions_meet_state_machines_an_institutional_approach}).

In this paper, we consider an IL which includes several distinct $loop$ operators and provide it with a denotational semantics, directly comparable to that given by \cite{UML_interactions_meet_state_machines_an_institutional_approach}. The semantics of an interaction with $loops$ is defined by considering any finite number of loop unfolding combinations. Then, we introduce a second semantics, which can be qualified as operational, as we aim at presenting it in the style advocated in \cite{Plotkin82}. Here, accepted traces of an interaction $i$ are defined by identifying its initial actions $act$, and for each of those the subsequent interaction $i'$ that will express the remainder of the trace. This operational semantics can therefore be thought of as a set of rules of the form \resizebox{!}{10pt}{$i \xrightarrow{act} i'$}. Doing so is however challenging as we need to keep track of possible conflicts between actions occurring on the same lifeline. While the operational semantics is particularly suitable to be adapted into concrete trace analysis algorithms, the denotational semantics serves as a mathematical foundation, revealing interesting algebraic properties. Both semantics have been implemented for semantic computation and conducted experiments indicate identical results. A trace analysis tool has also been adapted from the operational semantics and experimented on for correctness and performances.

The paper is organized as follows: Sec.\ref{sec:interaction_language_and_denotational_semantics} introduces the IL and the denotational semantics. Sec.\ref{sec:operational_semantics} and Sec.\ref{sec:trace_analysis} resp. introduce the operational semantics and the subsequent trace analysis algorithm while Sec.\ref{sec:comparison_of_both_approaches_for_semantics_generation} reports experimental results about the consistency of both semantics w.r.t. one another. Finally, Sec.\ref{sec:related-work} and Sec.\ref{sec:conclusion} resp. discuss related works and provide concluding remarks.

\section{Interaction language and denotational semantics\label{sec:interaction_language_and_denotational_semantics}}

\subsection{Base syntax\label{sec:basic_interaction_synt}}

This section provides a textual denotation of our basic IL (i.e. without loops). Interactions are defined up to a given signature $(L,M)$ where $L$ and $M$ resp. are sets of lifelines and messages. Their base building blocks are a set of communication actions (actions) over $L$ and $M$: $Act(L,M)=\{l \Delta m | l \in L, \Delta \in \{!,?\},m \in M\}$ where $l!m$ (resp. $l?m$) designates the emission (resp. reception) of the message $m$ from (resp. on) the lifeline $l$. For any action $act$ in $Act(L,M)$ of the form $l \Delta m$, $\theta(act)$ denotes the lifeline $l$. Actions can be composed using different binary operators that introduce an order of execution between them (weak or strict sequentiality, parallelism, mutual exclusivity).

\begin{definition}[Basic Interactions\label{def:basic_interaction}]
The set $\mathbb{B}(L,M)$ of basic interactions over $L$ and $M$ is inductively defined as follows:
\begin{itemize}
\item $\varnothing \in \mathbb{B}(L,M)$ and $Act(L,M) \subset \mathbb{B}(L,M)$,
\item $\forall (i_{1},i_{2}) \in \mathbb{B}(L,M)^{2}$ and $\forall f \in \{strict,seq,alt,par\}$, $f(i_{1},i_{2}) \in \mathbb{B}(L,M)$.
\end{itemize}
\end{definition}

The empty interaction $\varnothing$ and actions of $Act(L,M)$ are elementary interactions. The $strict$ and $seq$ operators are sequential operators: in $strict(i_1,i_2)$, all the actions in $i_1$ must take place before any action in $i_2$ while in $seq(i_1,i_2)$ sequentiality is only enforced between actions that share the same lifeline. In Fig.\ref{fig:umlsd_style}-b, $b?m$ may precede\footnote{Note that we omit depicting $seq$ on diagrams as is classically done in UML-SD.} $a!m$ (because $a \neq b$) while in Fig.\ref{fig:umlsd_style}-c $b?m$ cannot precedes $a!m$. 
Hence we use $strict$ to encode the emission and reception of the same message object e.g. $strict(a!m,b?m)$ on Fig.\ref{fig:umlsd_style}-c\footnote{drawn by convention as a plain arrow between $a$ and $b$}. In $alt(i_1,i_2)$, the behaviors specified by $i_1$ and $i_2$ are both acceptable albeit mutually exclusive\footnote{note that we handle the UML-SD $opt$ operator as $opt(i)=alt(i,\varnothing)=alt(\varnothing,i)$}. 
In Fig.\ref{fig:syntax_and_positions} if $a!m_1$ happens then $b?m_2$ cannot happen and vice-versa. In $par(i_1,i_2)$, the executions of $i_1$ and $i_2$ are interleaved. For instance, in $par(a!m_1,a!m_2)$, actions $a!m_1$ and $a!m_2$ can happen in any order.

Interactions being defined as usual terms, we use positions expressed in Dewey decimal notation to refer to subinteractions \cite{Dershowitz_rewrite_systems}. A position $p$ of $i$ is a sequence of positive integers denoting a path leading from the root node of $i$ to the subterm of $i$ at position $p$. 
Interactions are defined with operations whose arity is at most 2.
Hence, positions are words of $\{1,2\}^*$ i.e. words built over the empty word $\epsilon$, the words $1$ and $2$ and the concatenation law "$.$". 
In the following, we will use simplified notations without dots, e.g. "$11$" for the position "$1.1$".

In Def.\ref{def:positions_of_a_basic_interaction}, the functions $ST$ and $pos$ resp. associate to any interaction the set of all its subinteractions and the set of its positions. Moreover, we use the usual notation $i_{|p}$ \cite{Dershowitz_rewrite_systems} to designate unambiguously the subinteraction of $i$ at position $p$ for $p \in pos(i)$ (cf. example in Fig.\ref{fig:syntax_and_positions}).

\begin{definition}[Positions and subinteractions of a basic interaction\label{def:positions_of_a_basic_interaction}]
We define $ST : \mathbb{B}(L,M) \rightarrow \mathcal{P}(\mathbb{B}(L,M))$, $pos : \mathbb{B}(L,M) \rightarrow \mathcal{P}(\{1,2\}^*)$ and\footnote{ \_$_{|\_}$ is a partial function so that $i_{|p}$ is only defined  for positions occurring in $pos(i)$.} \_$_{|\_} : \mathbb{B}(L,M) \times \{1,2\}^* \rightarrow \mathbb{B}(L,M) $ such that $\forall i \in \mathbb{B}(L,M)$:
\begin{itemize}
\item if $i=\varnothing$ or $i \in Act(L,M)$ then $ST(i) = \{i\}$, $pos(i) = \{\epsilon\}$ and $i_{|\epsilon} = i$
\item if $i=f(i_1,i_2)$ with $f\in \{strict,seq,par,alt\}$ then:
    \begin{itemize}
        \item $ST(i) = \{i\}\cup ST(i_1) \cup ST(i_2)$ 
        \item $pos(i) = \{\epsilon\} \cup 1.pos(i_1) \cup 2.pos(i_2)$
        \item $i_{|\epsilon} = i$ and for $p=1.p'$ (resp. $2.p'$) in $pos(i)$, $i_{|p} = {i_1}_{|p'}$ (resp. ${i_2}_{|p'}$).
    \end{itemize}
\end{itemize}
\end{definition}

\subsection{Denotational semantics for basic interactions}

As explained in Sec.\ref{sec:basic_interaction_synt}, operators occurring in an interaction induce relations of precedence between the actions of the interaction. 
In the example of Fig.\ref{fig:syntax_and_positions}, if the left branch of the $alt$ is chosen (i.e. $a!m_1$ at position $11$) then the action $a!m_3$ at position $2$ must occur after it. However if the other branch were chosen (i.e. $b?m_2$ at position $12$), there would be no precedence order between actions $b?m_2$ and $a!m_3$ as their common ancestor is a $seq$ operator which only orders actions sharing the same lifeline. As a result, several orderings can be defined, depending, among others, on the choice of $alt$ branches. 
These possible orderings can be encoded as a set $ord(i)$ (defined in 
Def.\ref{def:ordering_basic_interaction}) which contains elements of the form $(e,o)$ where $e$ is the set of positions of the involved actions and $o$ reflects the precedence relations between those. In the example of Fig.\ref{fig:syntax_and_positions}, we have $ord(i) = \{ (\{11,2\}, \{(11,2)\}), (\{12,2\}, \emptyset)\}$. 
Indeed, as explained earlier, if the $11$ branch is chosen then the only two actions to be considered are $a!m_1$ and $a!m_3$ on resp. positions $11$ and $2$ (therefore $e=\{11,2\}$) and they are ordered because of both the $seq$ operator and their common lifeline, so that the associated precedence relation is modelled by $o = \{(11,2)\}$ meaning that $a!m_1$ at position $11$ should occur before $a!m_3$ at position $2$. 
The only other possible ordering occurs when branch $12$ is chosen and likewise we would have $e=\{12,2\}$ with $o=\emptyset$ because the $seq$ does not constrain the order of actions $b?m_2$ and $a!m_3$ with different lifelines.

\begin{definition}[Ordering type\label{def:ordering_type}]
Given $i$ in $\mathbb{B}(L,M)$. The set $\mathbb{O}(i)$ of candidate orderings of $i$ contains all couples $(e,o)$ such that \textbf{(1)} $e \subseteq pos(i)$, 
\textbf{(2)} for any $p$ in $e$, $i_{|p} \in Act(L,M)$
and \textbf{(3)} $o \subseteq e \times e$. $\mathbb{O}$ is then the set $\bigcup_{i \in \mathbb{B}(L,M)}\mathbb{O}(i)$.
\end{definition}

In Def.\ref{def:ordering_basic_interaction}, for a given interaction $i$, $ord(i)$ precisely defines which orderings are to be considered among the candidate orderings $\mathbb{O}(i)$. For an ordering $(e,o)$ in $\mathbb{O}$ and $p \in \{1,2\}$, we use the notation $p.e = \{p.p' | p' \in e\}$, $p.o = \{ (p.p_1,p.p_2) | (p_1,p_2) \in o\}$ and $p.(e,o) = (p.e,p.o)$. The notation is canonically extended to any set $O$ of orderings, by $p.O = \{p.(e,o)| (e,o) \in O\}$.

For the interaction $\varnothing$, there is no associated action and therefore we have a single $(e,o)=(\emptyset,\emptyset)$. For $a \in Act(L,M)$, there is a single action $a$ (at position $\epsilon$) and as a result, $ord(a)$ contains a single $(e,o)=(\{\epsilon\},\emptyset)$. For $i=alt(i_1,i_2)$, either $i_1$ or $i_2$ is executed. Thus any ordering in $ord(i)$ is simply an ordering from $ord(i_1)$ or from $ord(i_2)$ but correctly prefixed. 
Concretely, for any orderings $(e_1,o_1) \in ord(i_1)$ and $(e_2,o_2) \in ord(i_2)$, $ord(i)$ contains both $1.(e_1,o_1)$ and $2.(e_2,o_2)$. For $i=par(i_1,i_2)$, both $i_1$ and $i_2$ have to be executed but no order is enforced between actions of either child branch. Thus, for any ordering $(e_1,o_1) \in ord(i_1)$ and $(e_2,o_2) \in ord(i_2)$, $ord(i)$ contains $(1.e_1 \cup 2.e_2, 1.o_1 \cup 2.o_2)$. For $i=strict(i_1,i_2)$ both $i_1$ and $i_2$ have to be executed and all actions from $i_1$ must occur before actions from $i_2$. Thus for any orderings $(e_1,o_1) \in ord(i_1)$ and $(e_2,o_2) \in ord(i_2)$, $ord(i)$ contains an ordering $(e,o)$ that concerns all actions from both children i.e. $e=1.e_1\cup 2.e_2$ and such that $o$ keeps track of all initial precedence relations while incorporating those induced by the $strict$ operator i.e. $o=1.o_1\cup 2.o_2 \cup \{ (p_1 , p_2) | p_1 \in 1.e_1 , p_2 \in 2.e_2\}$. 
For $i=seq(i_1,i_2)$ the same reasoning can be applied, with the exception that additional precedence relations only concern actions that share the same lifelines. Using the same notations, $e=1.e_1\cup 2.e_2$ and $o=1.o_1\cup 2.o_2 \cup \{ (p_1 , p_2) | p_1 \in 1.e_1 , p_2 \in 2.e_2, \theta(i_{|p_1}) = \theta(i_{|p_2})\}$.

\begin{definition}[Orderings of a basic interaction\label{def:ordering_basic_interaction}]
We define the function $ord : \mathbb{B}(L,M) \rightarrow \mathcal{P}(\mathbb{O})$ as follows:
\begin{equation*}
\begin{array}{c}
ord(\varnothing) = \emptyset \quad \text{and} \quad
\forall \; act \in Act(L,M), \text{  } ord( act ) = \{ (\{\epsilon\}, \emptyset) \}
\end{array}
\end{equation*}
For any $i_1$ and $i_2$ in $\mathbb{B}(L,M)$:
\[
ord( alt(i_1,i_2) ) = 1.ord(i_1)  \cup 2.ord(i_2)
\]
\begin{equation*}
ord( par(i_1,i_2) ) = \bigcup_{\substack{(e_1,o_1) \in ord(i_1)\\(e_2,o_2) \in ord(i_2)}} \{ (1.e_1 \cup 2.e_2,1.o_1 \cup 2.o_2) \}
\end{equation*}
\begin{equation*}
ord( strict(i_1,i_2) ) = \bigcup_{\substack{(e_1,o_1) \in ord(i_1)\\(e_2,o_2) \in ord(i_2)}} \left\lbrace (e, o)  \middle| \begin{array}{l}
e=(1.e_1 \cup 2.e_2) \; , \; o=1.o_1 \cup 2.o_2 \cup o'\\
o' = \{ (p_1,p_2) \; | \;
    p_1 \in 1.e_1 \; , \; p_2 \in 2.e_2
    \} 
\end{array}
\right\rbrace
\end{equation*}
\begin{equation*}
ord( seq(i_1,i_2) ) = \bigcup_{\substack{(e_1,o_1) \in ord(i_1)\\(e_2,o_2) \in ord(i_2)}} \left\lbrace (e, o)  \middle| \begin{array}{l}
e=(1.e_1 \cup 2.e_2) \; , \; o=1.o_1 \cup 2.o_2 \cup o'\\
o' = \left\lbrace (p_1,p_2) \middle| 
    \begin{array}{l}
    p_1 \in 1.e_1 \; , \; p_2 \in 2.e_2 \\
    \theta(i_{|p_1}) = \theta(i_{|p_2})
    \end{array}
    \right\rbrace
\end{array}
\right\rbrace
\end{equation*}
\end{definition}

A given ordering $(e,o)$ with $e=\{e_1,...,e_n\}$ characterizes a set of behaviors that expresses every action whose position belongs to $e$ exactly once. 
Such a behavior is thus given under the form of an execution trace $i_{|e_{\alpha(1)}}...i_{|e_{\alpha(n)}}$ where $\alpha$ is a permutation of $[1,n]$. 
Obviously, not all of those permutations are acceptable as they must not contradict the partial order specified by $o$. If we note $p_j = e_{\alpha(j)}$ for $j$ in $[1,n]$, we have $\forall j,k\in [1,n]^2$ $j>k \Rightarrow (p_j,p_k) \not\in o$.

The semantics $\sigma(i)$ of an interaction $i$ then comes naturally as the union of all sets $sem(i,e,o)$ of execution traces of $i$ compatible with $(e,o) \in ord(i)$. When considering the example from Fig.\ref{fig:syntax_and_positions}, we have $sem(i,\{11,2\},\{(11,2)\}) = \{ a!m_1.a!m_3 \}$ and $sem(i,\{12,2\},\emptyset) = \{ b?m_2.a!m_3, a!m_3.b?m_2 \}$.

\begin{definition}[Denotational semantics for basic interactions\label{def:basic_interaction_semantics}]
For $i\in \mathbb{B}(L,M)$ and $(e,o)\in ord(i)$ with $n \in \mathbb{N}$ being the cardinal of $e$, we note:
\begin{equation*}
\begin{array}{l} 
sem(i,e,o) = \left\lbrace i_{|p_1}...i_{|p_n}  
\middle| \forall (p_j,p_k) \in e^2, \; j>k \Rightarrow p_j \neq p_k \land (p_j,p_k) \not\in o
\right\rbrace
\end{array}
\end{equation*}
$\sigma : \mathbb{B}(L,M) \rightarrow \mathcal{P}( Act(L,M)^* )$ is s. t. $\forall i \in \mathbb{B}(L,M)$,
$\sigma(i) = \underset{(e,o) \in ord(i)}{\bigcup} sem(i,e,o)$
\end{definition}

\subsection{Extension of the language with loops\label{sec:extension_with_loops}}

A loop is a repetition operator. Its content can be instantiated any finite number of times i.e multiple copies of it are inserted into the interaction. For UML-SD, the norm \cite{UMLNorm} states that "\textit{the loop construct represents a recursive application of the seq operator where the loop operand is sequenced after the result of earlier iterations}". 
The UML-SD loop is hence associated with the $seq$ operator. 
When instantiated, the loop content is ordered using $seq$ this means for example that $loop(a!m)$ becomes $seq(a!m,loop(a!m))$ then $seq(a!m,seq(a!m,loop(a!m)))$ and so on. In line with this explanation, let's consider the 4 types of loops that can be characterized according to the operator ordering the instantiated content ($seq$, $strict$, $par$ or $alt$). We can discard $alt$ as instantiating $loop(i)$ would lead to $alt(i,loop(i))$ meaning that the content can be read at most once and is therefore equivalent to $opt(i)$ (i.e. $alt(i,\varnothing)$). 
We will here consider 3 operators denoted $loop_{seq}$ (the classical loop), $loop_{strict}$ and $loop_{par}$.

\begin{figure}[h]
\vspace{-0.75cm}
    \centering
    \begin{tabular}{cccc}
    \includegraphics[width=0.25\textwidth]{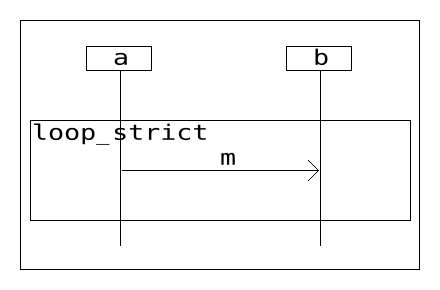} & \includegraphics[width=0.25\textwidth]{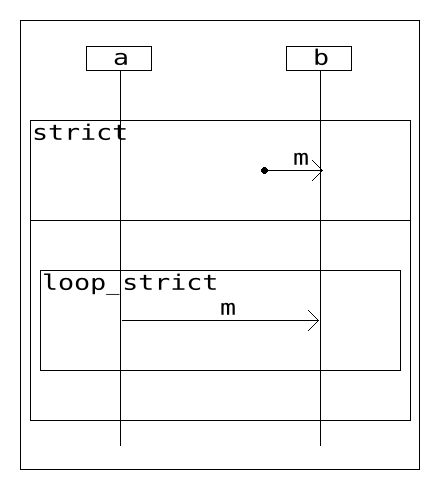} &
    \includegraphics[width=0.14\textwidth]{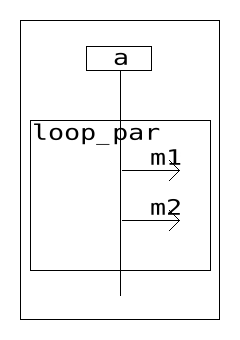} &
    \includegraphics[width=0.14\textwidth]{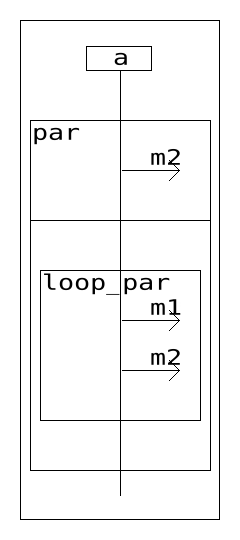}\\
    {\scriptsize(a-i) $i_a$} & {\scriptsize(a-ii) $i_a$ after $a!m$} & {\scriptsize(b-i) $i_{b}$} & {\scriptsize(b-ii) $i_b$ after $a!m_1$}
    \end{tabular}
    \caption{Examples showcasing the pertinence of $loop_{strict}$ and $loop_{par}$}
    \label{fig:example_loop_strict_par_pertinence}
\vspace{-0.75cm}
\end{figure}

In Fig.\ref{fig:example_loop_strict_par_pertinence}-a-i, $i_{a|11}=a!m$ is the only immediately executable action and its execution leads to the interaction $i'_a=strict(b?m,i_a)$ drawn on Fig.\ref{fig:example_loop_strict_par_pertinence}-a-ii. Because of the $strict$ operator, $i'_{a|211}=a!m$ is not immediately executable (preceded by $i'_{a|1}=b?m$). As a result $t_a=a!m.a!m.b?m.b?m$ is not an accepted trace for $i_a$. However, if there was a $seq$ operator instead of the $strict$, $i'_{a|211}$ would be immediately executable and $t_a$ an accepted trace.

Similarly, in Fig.\ref{fig:example_loop_strict_par_pertinence}-b-i, $i_{b|11}=a!m_1$ is the only immediately executable action and its execution leads to $i_b'=par(a!m_2,i_b)$ drawn on Fig.\ref{fig:example_loop_strict_par_pertinence}-b-ii. Because of the $par$ operator, $i'_{b|211}=a!m_1$ is immediately executable. As a result $t_b=a!m_1.a!m_1.a!m_2.a!m_2$ is an accepted trace for $i_b$. However, if there was a $seq$ instead of the $par$, $i'_{b|211}$ would not be immediately executable and $t_b$ not an accepted trace.

Consequently, considering $loop_{par}$ and $loop_{strict}$ in addition to the classic $loop_{seq}$ improves expressiveness. In rough terms, $loop_{par}$ always allows new instantiations as each instance is executed in parallel w.r.t each others and the loop itself. $loop_{strict}$ on the contrary does not allow new instantiations as long as the previous instance has not been entirely executed. The behavior of $loop_{seq}$ is somewhat in the middle, instantiations being allowed depending on the current structure of actions preceding and within the loop. 

In the following, we'll extend our IL to loops and adapt previous definitions (from $\mathbb{B}(L,M)$ to $\mathbb{I}(L,M)$). As in Def.\ref{def:interaction}, any time we do so, we will only define the missing cases concerning loop terms.

\begin{definition}[Interactions\label{def:interaction}]
The set $\mathbb{I}(L,M)$ of interactions over $L$ and $M$ is inductively defined as follows:
\begin{itemize}
\item $\varnothing \in \mathbb{I}(L,M)$ and $Act(L,M) \subset \mathbb{I}(L,M)$,
\item $\forall (i_{1},i_{2}) \in \mathbb{I}(L,M)^{2}$ and $\forall f \in \{strict,seq,alt,par\}$, $f(i_{1},i_{2}) \in \mathbb{I}(L,M)$,
\item $\forall i \in \mathbb{I}(L,M)$ and $\forall f\in \{strict,seq,par\}$, $loop_f(i)\in \mathbb{I}(L,M)$.
\end{itemize}

The functions $ST : \mathbb{I}(L,M) \rightarrow \mathcal{P}(\mathbb{I}(L,M))$, $pos : \mathbb{I}(L,M) \rightarrow \mathcal{P}(\{1,2\}^*)$ \\ and ~$\_ _{|\_} : \mathbb{I}(L,M) \times \{1,2\}^* \rightarrow \mathbb{I}(L,M) $ are defined by extending to loop terms the corresponding functions  of Def.\ref{def:positions_of_a_basic_interaction}: \\ 
For all $i$ in $\mathbb{I}(L,M)$ of the form  $loop_f(i')$ with $f\in \{strict,seq,par\}$:
    \begin{itemize}
        \item $ST(i) = \{i\}\cup ST(i')$ 
        \item $pos(i) = \{\epsilon\} \cup 1.pos(i')$,
        \item $i_{|\epsilon} = i$ and for $p=1.p'$ in $pos(i)$, $i_{|p} = i'_{|p'}$.
    \end{itemize}
\end{definition}

\begin{wrapfigure}{r}{0.25\textwidth}
\vspace{-2.25cm}
\centering
\begin{tabular}{c}
\includegraphics[width=0.25\textwidth]{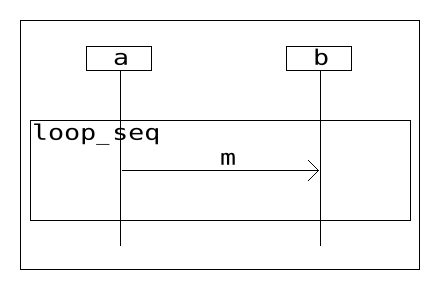}\\
{\scriptsize (a) $i=loop_{seq}(i_{|1})$}\\
{\scriptsize with $i_{|1}=strict(a!m,b?m)$}\\
\includegraphics[width=0.25\textwidth]{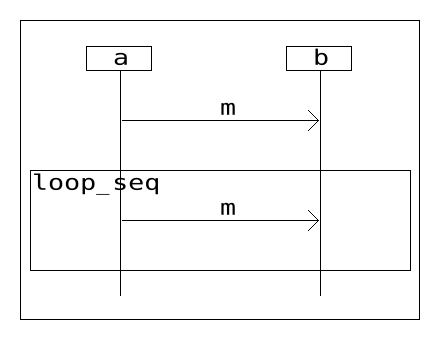}\\
{\scriptsize (b) $i'=seq(i_{|1},i)$}
\end{tabular}
\caption{Unfolding}
\label{fig:example_loop_unfold}
\vspace{-0.75cm}
\end{wrapfigure}

In order to define the semantics of interactions, we use the notion of term replacement \cite{Dershowitz_rewrite_systems}: the notation $t[s]_p$ denotes the term $t$ where its subterm at position $p$ is replaced by the term $s$. For instance with $i=seq(a!m,b?m)$, we have $i[c?m]_2=seq(a!m,c?m)$. This notation is convenient to represent terms obtained by loop unfolding. 
For example let us consider an interaction $i \in \mathbb{I}(L,M)$ with a $loop_{seq}$ at a position $p \in pos(i)$, that is, such that $i_{|p} = loop_{seq}(i_{|p.1})$. 
The interaction is then obtained from $i$ by unfolding once the loop at position $p$ is $i[seq(i_{|p.1},i_{|p})]_p$. 
In Def.\ref{def:n_unfoldings}, the set $\Upsilon(i,n)$ of all $n$-unfoldings of an interaction $i$ (i.e. the set of all interactions resulting from $n$ instantiations of \emph{any} loop from $i$) is defined recursively. On Fig.\ref{fig:example_loop_unfold} loop unfolding is illustrated with $\Upsilon(i,0)=\{i\}$ and $\Upsilon(i,1)=\{i'\}$.

\begin{definition}[$n$-unfoldings\label{def:n_unfoldings}]
We define $\Upsilon:\mathbb{I}(L,M) \times \mathbb{N} \rightarrow \mathcal{P}(\mathbb{I}(L,M))$ such that $\forall i \in \mathbb{I}(L,M)$ $\Upsilon(i,0) = \{i\}$ and $\forall n \in \mathbb{N}^+$:
\begin{equation*}
\Upsilon(i,n) =
\bigcup_{p \in pos(i) \text{ s.t. } i_{|p} = loop_f(i_{|p.1}) } \Upsilon( i[f(i_{|p.1},i_{|p})]_p, n-1)
\end{equation*}
\end{definition}

We define a function $F:\mathbb{I}(L,M) \rightarrow \mathbb{B}(L,M)$ that flattens interactions with loops i.e. that replaces all loop subterms with the empty interaction $\varnothing$. For instance, in Fig.\ref{fig:example_loop_unfold} we have $F(i)=\varnothing$ and $F(i')=seq(i_{|1},\varnothing)$. As $F(\mathbb{I}(L,M)) \subset \mathbb{B}(L,M)$, we can define an unfolding-based semantics\footnote{coined $\sigma_u$, $u$ standing for 'unfolding-based'} for $i \in \mathbb{I}(L,M)$ by simply considering the union of semantics obtained from flattened unfoldings of $i$.

\begin{definition}[Denotational semantics for interactions\label{def:trace_semantics_from_unfolding}]
\newline
We define $\sigma_u : \mathbb{I}(L,M) \rightarrow \mathcal{P}( Act(L,M)^* )$ such that for all $i$ in $\mathbb{I}(L,M)$:
\begin{equation*}
\sigma_u(i) = \bigcup_{n \in \mathbb{N}} \; \bigcup_{i' \in \Upsilon(i,n)}  \sigma(F(i'))
\end{equation*}
\end{definition}

\section{Operational Semantics\label{sec:operational_semantics}}

We aim to define algorithms that can determine whether or not a trace $t$ is accepted by an interaction $i$. 
This amounts to ascertaining whether or not $t \in \sigma_u(i)$. 
Naturally, being able to do so without having to compute $\sigma_u(i)$ is preferable. In the following we'll refer to this problem as 'trace analysis'.

\begin{wrapfigure}{l}{0.375\textwidth}
\vspace{-0.75cm}
\centering
\includegraphics[width=0.375\textwidth]{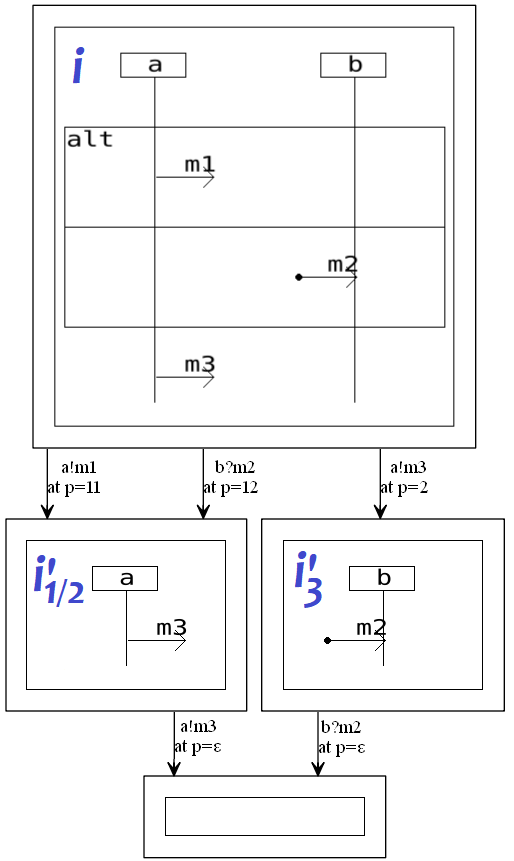}
\caption{Operational Semantics}
\label{fig:operational_semantics_applied}
\vspace{-0.75cm}
\end{wrapfigure}

As per Sec.\ref{sec:extension_with_loops}, asserting $t \in \sigma_u(i)$ equates to finding a combination of loop unfoldings $i^{\star} \in \bigcup_{k=0}^{\infty} \Upsilon(i,k)$ such that $t \in \sigma(F(i^{\star}))$. Even if feasible, this would be time and space consuming\footnote{and would not be adaptable if one considers an extension to monitoring as new combinations $i^{\star}$ may be needed every time a new action is observed}. As for non acceptation, it equates to proving that $\forall i^{\star} \in \bigcup_{k=0}^{\infty} \Upsilon(i,k)$ we have $t \not\in \sigma(F(i^{\star}))$. In this case, a termination in finite time would not even be guaranteed and would require defining some stopping criterion on the unfolding.

Consequently, we investigate another approach, in which traces are analyzed action by action. Here, instead of systematically unfolding loops, we do so on demand (when executing an $act$ that is found within a loop). This approach is based on a different semantics ($\sigma_o$) whose description is the purpose of Sec.\ref{sec:operational_semantics}. 

$\sigma_o$ is presented in the style of operational semantics, i.e. consisting in: \textbf{(1)} identifying from the structure of $i$ which $act$ can be immediately executed (coined 'frontier actions') and \textbf{(2)} deriving for each such $act$ a new interaction $i'$ specifying all the possible continuations of $act$ within the set of execution traces specified by $i$ (noted as \resizebox{!}{10pt}{$i \xrightarrow{act} i'$}). 

Intuitively, an action is in the frontier iff no structural operators (parent nodes) coerce it to be preceded by another action (sibling leaf). Accepted traces are then built recursively through the successive consumption of actions. Let's consider a trace $t=act_1.(...).act_n$ with $\forall k \in [1,n]$ \resizebox{!}{10pt}{$i_{k-1} \xrightarrow{act_k} i_k$} and such that $i_0=i$ (by extension we may note \resizebox{!}{10pt}{$i \xrightarrow{t} i_n$}).\\
\noindent $\bullet$ If the last interaction $i_n$ can express the empty trace $\epsilon$ (i.e. $\epsilon \in \sigma_u(i_n)$) - which can be statically analysed - then $t$ is accepted by $i$ i.e. $t \in \sigma_o(i)$.\\
\noindent $\bullet$ In any case, for all frontier actions $act_{n+1}$ of $i_n$, we have \resizebox{!}{10pt}{$i_n \xrightarrow{act_{n+1}} i_{n+1}$}, meaning that $t$ can be extended by $act_{n+1}$ and is a prefix of given trace(s) accepted by $i$.

To illustrate this, let's consider the example from Fig.\ref{fig:operational_semantics_applied}. The initial interaction is $i=seq(alt(a!m_1,b?m_2),a!m_3)$. There are 3 frontier actions that may play the role of $act$: $i_{|11}=a!m_1$, $i_{|12}=b?m_2$ and $i_{|2}=a!m_3$. 
The interactions remaining after the execution of $i_{|11}$ and $i_{|12}$ (resp. referred to as $i'_1$ and $i'_2$), which happen to be the same, are depicted below on the left, while the one remaining after the execution of $i_{|2}$ (noted $i'_3$) is depicted on the right.
The cases leading to $i'_1$ and $i'_2$ are self-evident. As for the one leading to $i'_3$, the execution of $a!m_3$ is contingent to the choice of the branch $12$ of the $alt$ hence the elimination of branch $11$ in the remaining interaction. Indeed, if branch $11$ were to be chosen, the execution of $a!m_3$ would not be possible as $a!m_1$ should have been executed before. This illustrates that $a!m_3$ is a frontier action up to the choice of the right branch of the $alt$ operator. Let us remark that $b?m_2$ may indeed happen after $a!m_3$ as those two actions occur on different lifelines and the top $seq$ operator structuring them does not constrain their order of execution. Finally, we conclude by defining the operational semantics as $\sigma_o(i) = a!m_1.\sigma_o(i'_1) \cup b?m_2.\sigma_o(i'_2) \cup a!m_3.\sigma_o(i'_3)$.

\subsection{Frontier actions}

In this section we explain how to identify frontier actions. Our notion of frontier differs slightly from that of \cite{tata2007}, where it refers to the set of positions $p$ such that $\forall j \in \{1,2\}^*$, $p.j \not\in pos(i)$ (i.e. positions of leaf nodes). Indeed, our frontiers contain only leaves that are immediately executable actions. 

Any ordering as defined in Def.\ref{def:ordering_basic_interaction} provides a partial order relation for the set of (positions of) actions of a basic interaction. A frontier action $act$ on position $p$ is then simply a minimal element given such a relation $(e,o)$, i.e. s.t. $\forall p' \in e$ we have $(p',p) \not\in o$ i.e. $act$ does not have to be preceded by any other action. 
The frontier of an interaction $i$ is then defined as the union of such $p$, considering all the orderings from $ord(i)$. As Def.\ref{def:ordering_basic_interaction} did not include $loop$ operators, we extend it in the following definition, in which the empty ordering $(\emptyset,\emptyset)$ corresponds to the case where the loop has not unfolded. According to this, the frontier of $i$ from Fig.\ref{fig:operational_semantics_applied} is then $front(i) = \{11,12,2\}$.

\begin{definition}[Ordering\label{def:ordering_interaction}]
We define $ord : \mathbb{I}(L,M) \rightarrow \mathcal{P}(\mathbb{O})$ as an extension to $\mathbb{I}(L,M)$ of its counterpart from Def.\ref{def:ordering_basic_interaction}. For all $f$ in $\{strict,seq,par\}$:
\begin{equation*}
\forall i \in \mathbb{I}(L,M) \text{,  } ord(loop_{f}(i)) = 1.ord(i) \cup  \{ (\emptyset,\emptyset) \}
\end{equation*}
\end{definition}

\begin{definition}[Frontier\label{def:frontier}]
$front : \mathbb{I}(L,M) \rightarrow \mathcal{P}(\{1,2\}^*)$ is the function s.t.:
\begin{equation*}
\forall i \in \mathbb{I}(L,M) \text{,  } front(i) = \bigcup_{(e,o) \in ord(i)} \{ p \in e \; | \; \forall p' \in e,\; (p',p) \not\in o \}
\end{equation*}
\end{definition}

\subsection{Pruning}

The design of the rules \resizebox{!}{10pt}{$i \xrightarrow{act} i'$} hinted at earlier is made operational thanks to 2 mechanisms: pruning and execution. Given an action $act\in front(i)$, branches preventing its execution are detected and eliminated with pruning. However, this is not done on the whole interaction $i$ but rather on specific neighboring (w.r.t. $act$) subinteractions. Execution orchestrates the calls to pruning, eliminates $act$ and constructs the remaining interaction $i'$.

\begin{figure}
\vspace{-0.5cm}
    \centering
\begin{tabular}{ccc}
\makecell{\includegraphics[width=0.25\textwidth]{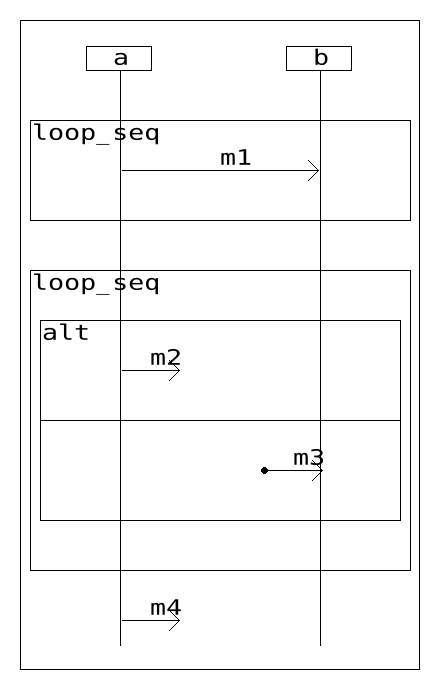}\\{\scriptsize(a) $i$}} & \makecell{\resizebox{0.3\textwidth}{!}{
            \begin{tikzpicture}[every node/.style = {shape=rectangle, align=center}]
	            \node (o) { $seq$ } [sibling distance=2.5cm,level distance=0.75cm]
            	    child {	node (o1) {	$loop_{seq}$ } [sibling distance=1.25cm]
	            	    child { node (o11) { $strict$ }
	            	        child { node (o111) {$a!m_1$ }}
            		    child { node (o112) {$b?m_{1}$}}
            			    }
	            	}
	                child { node (o2) { $seq$ } [sibling distance=1.25cm]
	                    child {	node (o21) {$loop_{seq}$}
	            	        child { node (o211) { $alt$ }
	            	            child { node (o2111) {$a!m_2$}}
            		            child { node (o2112) {$b?m_3$}}
            			        }
	            	        }
	                child { node (o22) {$a!m_4$ }}
	                };
	            \begin{pgfonlayer}{background}
\draw[green,fill=green,opacity=0.2](o1.north east) to[closed,curve through={(o11.east) .. (o112.north east) .. (o112.south east) .. ($(o112.south west)!0.5!(o111.south east)$) .. (o111.south west) .. (o111.north west) .. (o11.west).. (o1.north west)}](o1.north);
\draw[green,fill=green,opacity=0.2](o21.north east) to[closed,curve through={(o211.east) .. (o2112.north east) .. (o2112.south east) .. ($(o2112.south west)!0.5!(o2111.south east)$) .. (o2111.south west) .. (o2111.north west) .. (o211.west).. (o21.north west)}](o21.north);
\draw[red,fill=red,opacity=0.2](o22.north east) to[closed,curve through={(o22.east) .. (o22.south east) .. (o22.south) .. (o22.south west) .. (o22.west) .. (o22.north west)}](o22.north);
\end{pgfonlayer};
\draw[blue,very thick] (o2111.south west) -- (o2111.north east);
\draw[blue,very thick] (o2111.south east) -- (o2111.north west);
\draw[blue,very thick] (o211.east) -- (o211.west);
\draw[->,blue,very thick] ([xshift=-5pt,yshift=-5pt] o2112.north east) to [bend right=45] ([xshift=2.5pt] o211.east);
\draw[blue,very thick] (o111.south west) -- (o111.north east);
\draw[blue,very thick] (o111.south east) -- (o111.north west);
\draw[blue,very thick] (o11.south west) -- (o11.north east);
\draw[blue,very thick] (o11.south east) -- (o11.north west);
\draw[blue,very thick] (o1.south west) -- (o1.north east);
\draw[blue,very thick] (o1.south east) -- (o1.north west);
            \end{tikzpicture}
}\\{\scriptsize (b)}\\
{\scriptsize red - action to execute}\\
{\scriptsize green - neighbors to prune}\\
{\scriptsize blue - pruning}} & \makecell{\includegraphics[width=0.25\textwidth]{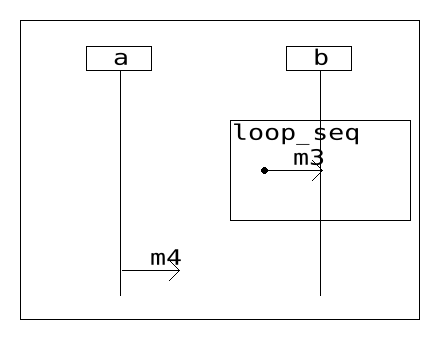}\\{\scriptsize (c) effect of pruning}\\\includegraphics[width=0.14\textwidth]{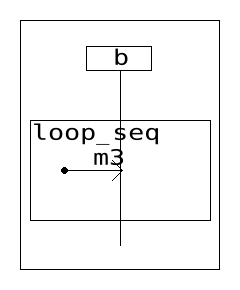}\\{\scriptsize (d) after executing $i_{|22}=a!m_4$}}
\end{tabular}
    \caption{Example showcasing pruning}
    \label{fig:example_loop_filtering}
\vspace{-0.5cm}
\end{figure}
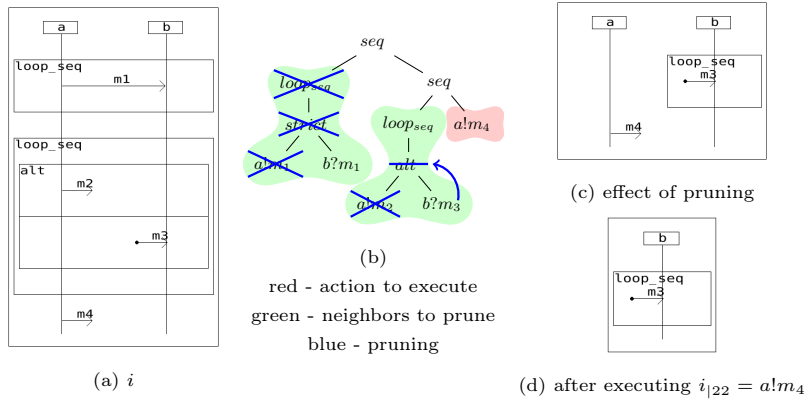

We first define the pruning mechanism which consists in removing from an interaction all the actions which occur on a given lifeline. For instance, on Fig.\ref{fig:example_loop_filtering}-b, let us consider the interactions $i_1=i_{|1}=loop_{seq}(strict(a!m_1,b?m_1))$ and $i_2=i_{|21}=loop_{seq}(alt(a!m_2,b?m_3))$ highlighted in green. We want to remove actions occurring on the lifeline $a$ (so as to allow the execution of $i_{|22} = a!m_4$). We find that $i_{1|11} = a!m_1$ (resp. $i_{2|11} = a!m_2$) needs to be removed from $i_1$ (resp. $i_2$).
If we do not want to get an interaction which is inconsistent or outwardly contradicts the original semantics, we can only prune subinteractions at positions where branching choices are made i.e. in $alt$ and $loop$ nodes. Indeed, by definition, eliminating a subinteraction at one such node would lead to a semantics that is included in the original.

In $i_2$, eliminating $i_{2|11}$ is easily done given that its parent node is an $alt$ and that its brother node does not need to be eliminated. Indeed, it suffices to operate the replacement $i_2[i_{2|12}]_1$ i.e. replacing the $alt$ node with its right child $b?m_3$. 

In $i_1$, eliminating $i_{1|11}$ is more delicate: its parent node is a $strict$ and as such, behaviors from its left and right children must both happen (there is no branching choice). Thus, if we want to eliminate $i_{1|11}$ we must also eliminate the whole $i_{1|1}$. The problem is hence forwarded upwards in the syntax. The parent $i_{1|\epsilon}$ is a loop operator, which characterizes a branching choice. We can eliminate the problematic branch by choosing not to instantiate the loop i.e. via the replacement $i_{1}[\varnothing]_\epsilon$.

The pruning mechanism is given in Def.\ref{def:pruning} as the recursive $prune$ function, which takes as arguments an interaction $i$ and a lifeline $l$. $prune$ eliminates from $i$ branching choices hosting actions that occur on $l$.

In a first descending phase, $prune$ goes down the syntax of $i$ through recursive calls (from root to leaves). When reaching a leaf, $prune$ returns an interaction $i'$ and a boolean $b$. $b=\top$ signifies that the current branch needs to be eliminated (pruned) while $i'$ is the interaction that will be used to reconstruct $i$ in the ascending phase (only used if $b=\bot$). 
Leaves are either actions or empty interactions. For an action $act$, if $\theta(act) = l$, the current branch must be pruned so $prune(act,l)=(\varnothing,\top)$: the value of the returned interaction $i'$ has no importance here because a parent will be pruned anyway. If $\theta(act) \neq l$ we have $prune(act,l)=(act,\bot)$ because there is nothing to prune here. Similarly, we have $prune(\varnothing,l)=(\varnothing,\bot)$. 

In the second, ascending phase, the pruned interaction is reconstructed according to the values of $i'$ and $b$ returned from child branches. If at any point $b = \top$, this value is forwarded upwards until an expendable branching choice is reached.

$prune(i,l)$ is recursively called on the child nodes of $i$. Depending on the operator in $i$, the return values of $prune(i_{|1},l) = (i'_1,b_1)$ (and also $prune(i_{|2},l) = (i'_2,b_2)$ for binary operators) will be used differently to determine $i'$ and $b$.

For the operators $f\in \{strict,seq,par\}$, if any one child must be pruned ($b_1 \vee b_2)$ then the whole branch must also be pruned and otherwise a reconstructed $f(i'_1,i'_2)$ is returned. For the exclusive alternative $alt$, if no branch needs pruning, $alt(i'_1,i'_2)$ is returned; if any single branch needs pruning, $prune$ returns the one that does not need to be pruned and if both branches need pruning, then the whole interaction is pruned. For the repetition operators, if the loop content needs pruning then the choice of 'never taking the loop' is made meaning that $\varnothing$ is returned with $b=\bot$, signifying a successful pruning. If there is no needed pruning, it simply returns the loop with an already pruned loop content $loop_f(i'_1)$.

\begin{definition}[Pruning\label{def:pruning}]
$prune : \mathbb{I}(L,M) \times L \rightarrow \mathbb{I}(L,M) \times bool$ is the function such that for all $i \in \mathbb{I}(L,M)$ and $l \in L$:
\begin{itemize}
    \item $prune(\varnothing,l) = (\varnothing,\bot)$
    \item for $act \in Act(L,M)$: if $\theta(act) = l$ then $prune(act,l) = (\varnothing,\top)$ (else $(act,\bot)$)
    \item if $i = f(i_1,i_2)$ with $f \in \{strict,seq,par\}$, given $prune(i_1, l) = (i_1',b_1)$ and $prune(i_2, l) = (i_2',b_2)$:\\
    if $b_1 \vee b_2$ then $prune(i,l) = (\varnothing, \top)$ (else $(f(i_1',i_2'), \bot)$)
    \item if $i=alt(i_1,i_2)$, given $prune(i_1,l) = (i_1',b_1)$ and $prune(i_2,l) = (i_2',b_2)$:
        \begin{itemize}
            \item if $b_1 \wedge b_2$ then $prune(i,l) = (\varnothing,\top)$
            \item if $b_1 \wedge \neg b_2$ then $prune(i,l) = (i_2',\bot)$
            \item if $\neg b_1 \wedge b_2$ then $prune(i,l) = (i_1',\bot)$
            \item if $\neg b_1 \wedge \neg b_2$ then $prune(i,l) = (alt(i_1',i_2'),\bot)$
        \end{itemize}
    \item if $i=loop_{f}(i_1)$ with $f \in \{strict,seq,par\}$, given $prune(i_1,l) = (i_1',b_1)$:\\
    if $b_1$ then $prune(i,l) = (\varnothing,\bot)$ (else $(loop_{f}(i_1'),\bot)$)
\end{itemize}
\end{definition}

\subsection{Execute function and operational semantics}

Let us consider the example $i$ from Fig.\ref{fig:example_loop_filtering}. 
We wish to execute the frontier action $i_{|22} = a!m_4$ (highlighted in red). 
To allow this execution we need at first to remove the actions occurring on the same lifeline (i.e. on $a$) from the neighbors highlighted in green. 
To do so, we use the $prune$ function from Def.\ref{def:pruning}. More generally, the nature of our syntax is such that, for the execution of a frontier action at position $p$, we only need to prune subinteractions at positions $p_{0}.1$ s.t. $\exists p' \in \{1,2\}^*$ s.t. $p = p_{0}.2.p'$ and s.t. $i_{|p_{0}} = seq(i_{|p_{0}.1},i_{|p_{0}.2})$. Those are exactly the left cousins of $i_{|p}$ that are scheduled sequentially (i.e. with $seq$) w.r.t. $i_{|p}$.

We now define the execution function $\chi$ (Def.\ref{def:execute_function}), which takes as arguments an interaction $i$ and a frontier position $p$ and returns the remaining interaction $i'$. As explained earlier, $\chi$ orchestrates the use of $prune$. In the example from Fig.\ref{fig:example_loop_filtering} this first cleaning feature would result in the transformation of $i$ from the diagram on Fig.\ref{fig:example_loop_filtering}-a to the one on Fig.\ref{fig:example_loop_filtering}-c. The only thing left to do is then to remove the executed action s.t. the result is the interaction from Fig.\ref{fig:example_loop_filtering}-d.

$\chi$ is defined inductively on both the structure of the interaction $i$ and the position $p=d_1...d_n \in \{1,2\}^n$. The execution of $\chi(i,p)$ traverses recursively the syntactic structure of $i$ guided by the path defined by the position $p$, that is, from $\chi(i_{|\epsilon},d_1...d_n)$ (root node), ..., up to $\chi(i_{|p},\epsilon)$ (target action leaf to execute).
Here, $\chi(i_{|p},\epsilon) = \varnothing$ constitutes the stopping criterion and $i'$ is then constructed when the algorithm goes back up through the syntactic structure of $i$. Assigning $\varnothing$ to $\chi(i_{|p},\epsilon)$ ensures that the action $i_{|p}$ is removed in the construction of $i'$.

When a $par$ node is encountered during the upward traversal, i.e. for $j\in [1,n]$,  $i_{|d_1...d_j}=par(i_{|d_1...d_j.1},i_{|d_1...d_j.2})$ then $\chi(i_{|d_1...d_j},d_{j+1}...d_n)$ is simply:

$par(\chi(i_{|d_1...d_j.1},d_{j+2}...d_n), i_{|d_1...d_j.2})$ if $d_{j+1}=1$ or,

$par(i_{|d_1...d_j.1}, \chi(i_{|d_1...d_j.2},d_{j+2}...d_n))$ if $d_{j+1}=2$. 

\noindent
Indeed, as $par$ specifies parallel executions, there is no need for pruning.

When an $alt$ node is reached, using the same notations, we would have:

$\chi(i_{|d_1...d_j},d_{j+1}...d_n) = \chi(i_{|d_1...d_{j+1}},d_{j+2}...d_n)$. 

\noindent
Indeed, we can 'skip' the $alt$ node itself and replace it directly with the interaction resulting from the execution of the chosen branch.

When a $loop$ is reached, i.e. $i_{|d_1...d_j}=loop_f(i_{|d_1...d_j.1})$ (with a mandatory $d_{j+1}=1$), we have :

$\chi(i_{|d_1...d_j},d_{j+1}...d_n) = f(\chi(i_{|d_1...d_{j+1}},d_{j+2}...d_n),i_{|d_1...d_j})$. 

\noindent
Indeed, the execution is done on a copy of the loop content that precedes (with $f$ operator) the loop $i_{|d_1...d_j}$ itself, that is, on an unfolding of the loop.

For the sequential operators, pruning needs to be considered only if the executing action is situated on the right branch of the $seq$ or $strict$ node (if the action is on the left branch, we have the same transformation as in the $par$ case). Given $i_{|d_1...d_j}=seq(i_{|d_1...d_j.1},i_{|d_1...d_j.2})$ and $d_{j+1}=2$, when constructing $\chi(i_{|d_1...d_j},d_{j+1}...d_n)$ we must prune in $i_{|d_1...d_j.1}$ all the actions that could interfere with $i_{|p}$ i.e. those taking place on $\theta(i_{|p})$. As such, given $(i'_1,b_1) = prune(i_{|d_1...d_j.1},\theta(i_{|p}))$, we'll replace the left branch of the $seq$ with $i_1'$ and reconstruct:

$\chi(i_{|d_1...d_j},d_{j+1}...d_n) = seq(i_1',\chi(i_{|d_1...d_{j+1}},d_{j+2}...d_n))$.

Given that the $strict$ operator won't allow any action from the left branch to occur after an action on the right has occurred, we can simply prune the whole left branch i.e. given $i_{|d_1...d_j}=strict(i_{|d_1...d_j.1},i_{|d_1...d_j.2})$ and $d_{j+1}=2$:

$\chi(i_{|d_1...d_j},d_{j+1}...d_n) = \chi(i_{|d_1...d_{j+1}}, d_{j+2}...d_n)$.

\begin{definition}[Execution\label{def:execute_function}]
The function $\chi : \mathbb{I}(L,M) \times \{1,2\}^* \rightarrow \mathbb{I}(L,M)$ is defined for couples $(i,p)$ with $i \in \mathbb{I}(L,M)$ and $p \in front(i)$ as follows:
\begin{itemize}
    \item if $p=\epsilon$ then $\chi(i,p)=\varnothing$
    \item if $p=1.p_1$ then
        \begin{itemize}
            \item if $i=f(i_1,i_2)$ with $f \in \{strict,seq,par\}$ then $\chi(i,p)=f(  \chi(i_1,p_1)  ,i_2)$
            \item if $i=alt(i_1,i_2)$ then $\chi(i,p)=\chi(i_1,p_1)$
            \item if $i=loop_{f}(i_1)$ with $f \in \{strict,seq,par\}$ then $\chi(i,p)=f( \chi(i_1,p_1) ,i)$
        \end{itemize}
    \item if $p=2.p_2$ then
        \begin{itemize}
            \item if $i=seq(i_1,i_2)$ then $\chi(i,p)=seq(i'_1,\chi(i_2,p_2) )$ \\ where $ prune(i_{1},\theta(i_{|p})) = (i'_1,b)$
            \item if $i=strict(i_1,i_2)$ then $\chi(i,p)=\chi(i_2,p_2)$
            \item if $i=par(i_1,i_2)$ then $\chi(i,p)=par( i_1 , \chi(i_2,p_2) )$
            \item if $i=alt(i_1,i_2)$ then $\chi(i,p)=\chi(i_2,p_2)$
        \end{itemize}
\end{itemize}
\end{definition}

In Def.\ref{def:interaction_semantics_by_rewriting} below, we now define the operational semantics. Note that interactions that can express the empty trace $\epsilon$ are identified with the predicate $exp_{\epsilon}$. This semantics expresses rules of the form \resizebox{!}{10pt}{$i \xrightarrow{ i_{|p} } \chi(i,p)$} where $p \in front(i)$.

\begin{definition}[Operational semantics for interactions\label{def:interaction_semantics_by_rewriting}]
\newline
We define $\sigma_o : \mathbb{I}(L,M) \rightarrow \mathcal{P}( Act(L,M)^* )$ as:
\[
\sigma_o(i) = empty(i) \cup \bigcup_{p \in front(i)} i_{|p}.\sigma_o( \chi(i,p) )
\]
with $empty(i) = \{\epsilon\}$ (resp.$\emptyset$) if $exp_{\epsilon}(i) = \top$ (resp. $\bot$)    

\noindent
where $exp_{\epsilon} : \mathbb{I}(L,M) \rightarrow bool$ is defined as:
\begin{itemize}
\item $exp_{\epsilon}(\varnothing) = \top$
\item $exp_{\epsilon}(l\Delta m) = \bot$
\item $exp_{\epsilon}(f(i_1,i_2)) = exp_{\epsilon}(i_1) \wedge exp_{\epsilon}(i_2)$ for $f\in \{strict,seq,par\}$
\item $exp_{\epsilon}(alt(i_1,i_2)) = exp_{\epsilon}(i_1) \vee exp_{\epsilon}(i_2)$
\item $exp_{\epsilon}(loop_f(i_1)) = \top$ for $f\in \{strict,seq,par\}$
\end{itemize}
\end{definition}

\section{Back-to-back comparison of both semantics\label{sec:comparison_of_both_approaches_for_semantics_generation}}

\textbf{Dataset.} The recursive definition of interactions as syntactic terms allows to characterize them by their depth. Interactions of depth $1$ include the empty interaction $\varnothing$ and all actions from $Act(L,M)$. Depending on the cardinals $n_l = Card(L)$ and $n_m = Card(M)$, those interactions can all be enumerated and computed. Given a signature, interactions of depth $2$ can be deduced from those of depth $1$ and exhaustively computed via the application of the binary and unary operators (e.g. $seq(\varnothing ,a!m)$). Likewise, interactions of depth $3$ can be computed from those of depths $1$ and $2$ and so on. To illustrate this, Fig.\ref{fig:table_number_of_interactions} presents for each couple $(n_l,n_m)$ the numbers of interactions of depths $1$, $2$ and $3$ in each cell. For instance, we have $3$ interactions of depth $1$ for $n_l=n_m=1$.

\begin{wrapfigure}{r}{0.35\textwidth}
    \centering
    \resizebox{0.35\textwidth}{!}{
        \begin{tabular}{|c|c|c|c|}
        \hline
        \diaghead{\hskip1.4cm}%
        {$n_m$}{$n_l$} & 1         & 2         & 3 \\ \hline
        1                                           &    \makecell{3\\45\\9315}       & \makecell{5\\115\\57845}          & \makecell{7\\217\\201159} \\ \hline
        2                                           &   \makecell{5\\115\\57845}        & \makecell{9\\351\\519129}          & \makecell{13\\715\\2121405}  \\ \hline
        3                                           & \makecell{7\\217\\201159}          &  \makecell{13\\715\\2121405 }         & \makecell{19\\1501\\9244659} \\ \hline
        \end{tabular}
    }
    \caption{Numbers of interactions per $n_l$, $n_m$ and $d$}
    \label{fig:table_number_of_interactions}
\vspace{-1cm}
\end{wrapfigure}

\noindent \textbf{Experiments.} We implemented both semantics ($\sigma_u$ from Def.\ref{def:trace_semantics_from_unfolding} and $\sigma_o$ from Def.\ref{def:interaction_semantics_by_rewriting}) and compared the set of traces $\sigma_u(i)$ and $\sigma_o(i)$ they generate (with a stopping criterion on the maximum number of loop unfolding - $4$ in our experiments) on a significant set of interactions of depth $3$ with $n_l=n_m=3$.
For all of the $234175$ selected interactions $i$ from our dataset, the tests systematically concluded on the equality $\sigma_u(i)=\sigma_o(i)$. Although not a proof, our successful back-to-back comparison comforts our confidence in both semantics, all the more so because of the exhaustivity of the subject data set up to maximum numbers of lifelines, messages types, interaction depth (up to 3), number of loop unfolding (up to $4$), allowing covering all 2 by 2 combinations of operators.

\section{Trace analysis\label{sec:trace_analysis}}

\begin{wrapfigure}{r}{0.275\textwidth}
\vspace{-2cm}
    \centering
\includegraphics[width=0.275\textwidth]{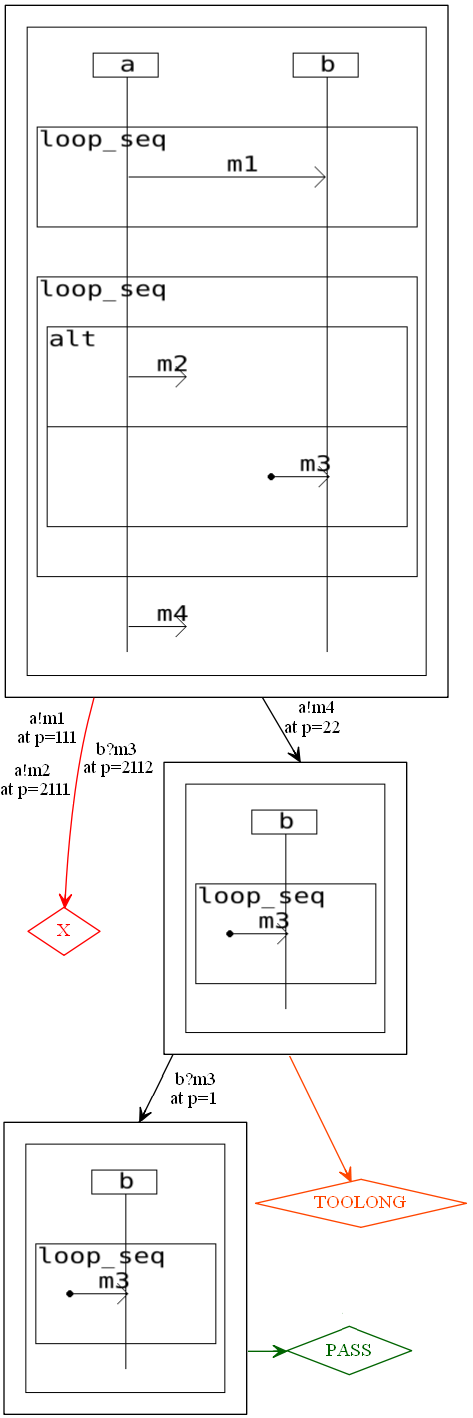}
    \caption{Application of $\omega$}
    \label{fig:application_trace_analysis}
\vspace{-1cm}
\end{wrapfigure}

The definition of the execution function $\chi$ (Def.\ref{def:execute_function}) that comes with the operational nature of the $\sigma_o$ semantics (Def.\ref{def:interaction_semantics_by_rewriting}) allows us to solve the 'trace analysis' problem hinted at earlier. Indeed, analysing a trace $t=act_1...act_n$ w.r.t. an interaction $i_0$ equates to verifying whether or not there exists transformations \resizebox{!}{10pt}{$i_0 \xrightarrow{act_1} \chi(i_0,p_1) = i_1$}, ..., \resizebox{!}{10pt}{$i_{n-1} \xrightarrow{act_n} \chi(i_{n-1},p_n) = i_n$} s.t. $i_n$ accepts the empty trace.

We define an $\omega$ function (Def.\ref{def:trace_analysis}) which takes as arguments an interaction $i$ and a trace $t$ and checks whether or not $t$ is a trace of $i$. Additional traceability information is provided using four distinct verdicts:

\noindent $\bullet$ $Covered$ is returned when $t$ is a trace of $i$ i.e. $t \in \sigma_o(i)$;

\noindent $\bullet$ $TooShort$ is returned when $t \not\in \sigma_o(i)$ is a strict prefix of a trace of $i$ i.e. $\exists t' \in Act(L,M)^*$ s.t. $t.t' \in \sigma_o(i)$;

\noindent $\bullet$ $TooLong$ is returned when neither $Covered$ nor $TooShort$ can be, and given $t=act_1...act_n$ $\exists k < n$ s.t. $act_1...act_k \in \sigma_o(i)$ i.e. $t$ extends a trace of $i$;

\noindent $\bullet$ $Out$ is returned when none of the others can be.

We define the enumerated type $Verdict$ and provide it with a total order $Out \prec TooLong \prec TooShort \prec Covered$.

\noindent $\bullet$ If $t$ is empty then: either $i$ accepts the empty trace in its semantics and in this case  $\omega(i,t)$ returns $Covered$, or it returns $TooShort$.

\noindent $\bullet$ If $t$ is of the form $act.t'$ (i.e. not empty and starts with $act$) then, for all matching actions $i_{|p}$ in the frontier of $i$, recursive calls are performed on $\omega(\chi(i,p),t')$ and $\omega(i,t)$ returns the strongest ($max_\prec$ function) verdict among those and either $TooLong$ if $i$ expresses the empty trace $\epsilon$ or $Out$ if not.

\begin{definition}[Trace Analysis\label{def:trace_analysis}]
We define $\omega : \mathbb{I}(L,M) \times Act(L,M)^* \rightarrow Verdict$ such that $\forall i,t \in \mathbb{I}(L,M) \times Act(L,M)^*$:
\begin{itemize}
\item $\omega(i,\epsilon) = Covered$ (resp. $TooShort$) if $exp_{\epsilon}(i) = \top$ (resp. $\bot$)
\item if $t$ is of the form $act.t'$ then:
\begin{equation*}
    \omega(i,t) = max_\prec \left( out_\epsilon(i) \cup  \left\{ \omega(\chi(i,p),t') \middle| \begin{array}{c}p \in front(i)\\i_{|p} = act\end{array} \right\} \right)
\end{equation*}
\end{itemize}
with $out_\epsilon(i) = \{TooLong\}$ (resp. $\{Out\}$) if $exp_{\epsilon}(i) = \top$ (resp. $\bot$)
\end{definition}

Fig.\ref{fig:application_trace_analysis} is a graphical representation of the $\omega$ process when applied to the interaction from Fig.\ref{fig:example_loop_filtering}-a and the trace $a!m_4.b?m_3$.

\begin{wrapfigure}{l}{0.275\textwidth}
\vspace{-0.75cm}
    \centering
    \resizebox{.275\textwidth}{!}{\rotatebox{90}{
\begin{tabular}{|l|l|l|l|l|l|l|l|l|l|l|}
\hline
\multicolumn{2}{|l|}{\multirow{4}{*}{Total 156276}} & trc & \multicolumn{4}{l|}{act} & prf & add & \multicolumn{2}{l|}{rep} \\ \cline{3-11}\multicolumn{2}{|l|}{} & 3231 & \multicolumn{4}{l|}{18000} & 4618 & 50600& \multicolumn{2}{l|}{79827}   \\ \cline{3-11} 
\multicolumn{2}{|l|}{} & cov & cov & short & long & out & short & \makecell{short\\or\\long} & \makecell{short\\or\\long} & \makecell{out\\or\\short} \\ \cline{3-11} 
\multicolumn{2}{|l|}{} & 3231 & 352 & 1705 & 864 & 15079 & 4618 & 50600 & 10948 & 68879 \\ \hline
COV & 3583 & 3231 & 352 & 0 & 0 & 0 & 0 & 0 & 0 & 0 \\ \hline
SHORT & 9927 & 0 & 0 & 1705 & 0 & 0 & 4618 & 358 & 505 & 2741 \\ \hline
LONG & 61549 & 0 & 0 & 0 & 864 & 0 & 0 & 50242 & 10443 & 0 \\ \hline
OUT & 81217 & 0 & 0 & 0 & 0 & 15079 & 0 & 0 & 0 & 66138 \\ \hline
\end{tabular}
    }}
    \caption{Correctness of $\omega$ experiments}
    \label{fig:experimental_results_correctness_trace_analysis}
\vspace{-0.75cm}
\end{wrapfigure}

Fig.\ref{fig:experimental_results_correctness_trace_analysis} presents a synthesis of experiments conducted to assess the correctness of $\omega$ and of our implementation of it. We randomly sampled $1000$ interactions from the set of $234175$ interactions mentioned in Sec.\ref{sec:comparison_of_both_approaches_for_semantics_generation}. Each of them were tested with the $18$ single action traces from $Act(L,M)$ and we sampled $15$ traces from their semantics (computed with $3$ loop unfolds). Each of those traces were tested as well as a random selection of their prefixes and of interesting mutants. Addition (resp. replacement) mutants consists in adding an action to a trace (resp. prefix). By construction we could classify all those traces according to the verdicts they are expected to obtain. Fig.\ref{fig:experimental_results_correctness_trace_analysis} details those results, showing a systematic concordance between the expected and obtained verdicts. Those results reinforce our confidence on $\omega$, the more so that they were done on a panel of traces and interactions which covers all 2 by 2 combinations of operators.

\begin{wrapfigure}{r}{0.55\textwidth}
\begin{tabular}{cc}
\makecell{\includegraphics[width=0.2\textwidth]{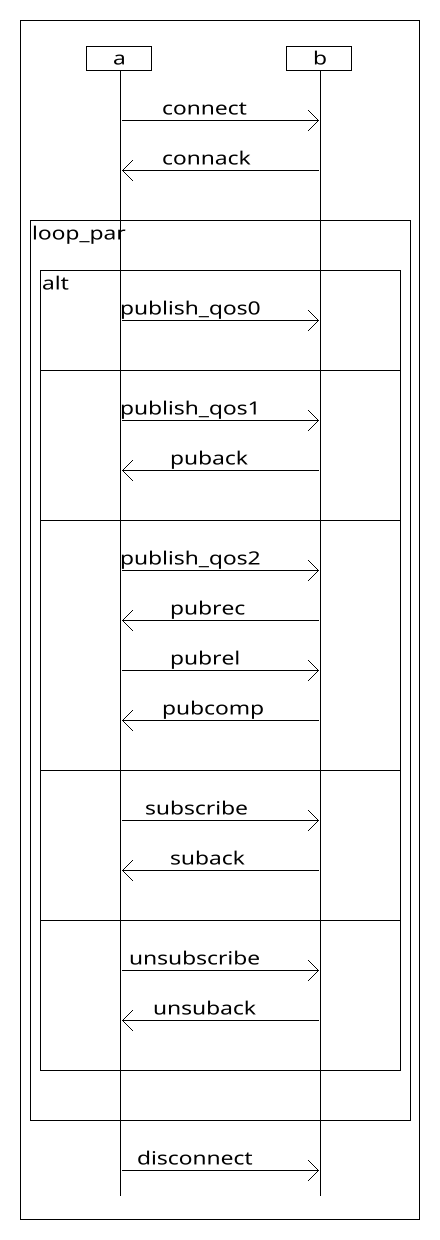}\\{\scriptsize (a) mqtt model}} & \makecell{\includegraphics[width=0.35\textwidth]{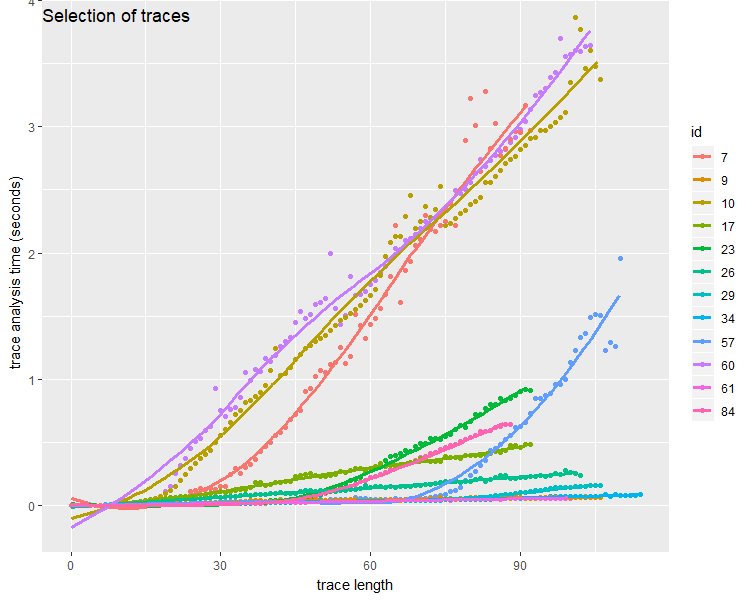}\\
\includegraphics[width=0.35\textwidth]{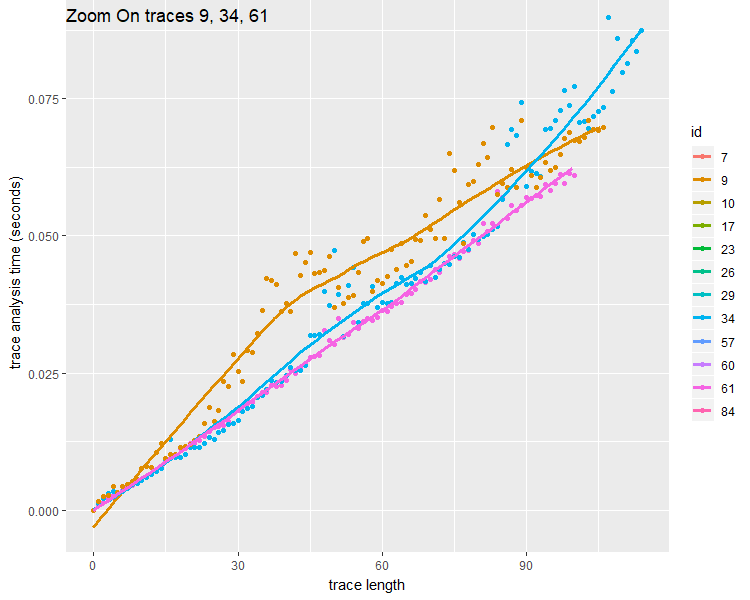}\\{\scriptsize (b) time vs. trace length}}
\end{tabular}
\vspace{-0.25cm}
\caption{Performances}
\label{fig:trace_analysis_performances}
\vspace{-0.5cm}
\end{wrapfigure}

To provide an evaluation of performances (plotting time vs. length), we needed a large model and long correct traces. Indeed, the time required by the analysis is not always correlated to trace length e.g. an arbitrarily long trace starting with an action $act$ of position $p \not\in front(i)$ is analyzed immediately, whatever length it may be. There is however a correlation for correct traces and their prefixes. We defined a partial high-level model of the MQTT \cite{MQTTNorm} telecommunication protocol (see Fig.\ref{fig:trace_analysis_performances}-a). This model states that a communication session between a client and a broker starts (resp. ends) with a sequential connection (resp. disconnection) phase. In between, at any time, any number of instances of one of the $5$ proposed subinteractions can be run concurrently. Hence, we used a multi-threaded Python script to generate $100$ traces, each of those corresponding to the concurrent activation and execution at random time intervals of $20$ instances of the $loop_{par}$ from Fig.\ref{fig:trace_analysis_performances}-a. All those traces (resp. prefixes) have the verdict $Covered$ (resp. $TooShort$); we evaluated computation times and plotted some of them on Fig.\ref{fig:trace_analysis_performances}-b. 

The linear regression shows curves with a great variability (some traces need $4$ seconds while others only $0.06$). In this precise model, it is explained by the presence of $par$ (via $loop_{par}$) operators and by the fact that messages are not uniquely identified. For instance analyzing $t=a!m.b?m$ on $i=par(a!m,strict(a!m,b?m))$ would give rise to $2$ branches: $i'=strict(a!m,b?m)$ (resp. $i'=par(a!m,b?m)$) with $t'=b?m$ which ends with $Out$ (resp. $Covered$) because $m$ is not uniquely identified. This number of branches can quickly explode when $par$ operators are stacked which happens when the trace describes an execution where many loop content instances overlap. An applicable solution is to treat message data arguments, given that communication protocols provide unique ids e.g. $m(id1)\neq m(id2)$. In Fig.\ref{fig:trace_analysis_performances}-b, on the plot below, we magnified on traces 9, 34 \& 61 which have a very short analysis time. We can surmise here that minimal (perhaps no) loop overlap occurred as the derivatives are almost constants (especially for trace 61). In conclusion, performance highly depends on the model and input trace, but treating data which specifies unique ids for messages would generalize the best case scenario. In this case, the algorithm could be applied to monitoring within the limits of an input frequency that is inferior to the time required to analyze a trace of length $1$.

\section{Related work\label{sec:related-work}}

For classical IL such as UML-SD or HMSC, many authors have proposed their own takes on formal semantics (see the survey \cite{the_many_meanings_of_uml2_sd_a_survey} for UML-SD).

\noindent \textbf{Denotational Semantics.} 
Most existing semantics based on term interpretations are given in a denotational style \cite{semantics_of_interactions_in_uml_2_0,STAIRS_towards_formal_design_with_sequence_diagrams,an_institution_for_uml_2_0_interactions,UML_interactions_meet_state_machines_an_institutional_approach} and do not follow-up with algorithmic tools.
In \cite{semantics_of_interactions_in_uml_2_0}, the authors propose a denotational semantics similar to ours (Def.\ref{def:basic_interaction_semantics}) as far as the $strict$, $alt$ and $par$ operators are concerned.
\cite{STAIRS_towards_formal_design_with_sequence_diagrams} proposes a semantics that is a detailed version of the one from \cite{semantics_of_interactions_in_uml_2_0}. In  \cite{UML_interactions_meet_state_machines_an_institutional_approach} there is a distinction ($snd(s,r,m)|snd(s,m)|rcv(s,r,m)|rcv(r,m)$) between basic actions whether or not the intended receiver or original sender is the environment. Apart from that, and the absence of $loops$, the denotational semantics proposed by \cite{UML_interactions_meet_state_machines_an_institutional_approach} is similar to ours. 
In \cite{an_institution_for_uml_2_0_interactions}, an institutional approach, likened to that of \cite{UML_interactions_meet_state_machines_an_institutional_approach} is proposed. However it includes $loops$ and deals with modalities associated to the $neg$ and $assert$ operators \cite{UMLNorm} by separating the semantics in sets of accepted and refused traces. This issue of modality is also raised in \cite{the_many_meanings_of_uml2_sd_a_survey} and \cite{assert_and_negate_revisited_modal_semantics_for_UML_sequence_diagrams} but it is out of the scope of this paper.

\noindent \textbf{Translations based approaches.}
Most other approaches rely on translations that map concepts of the given IL into a target formal framework, most often based on automata
\cite{timed_sequence_diagrams_and_tool_based_analysis_a_case_study,eliciting_unitary_constraints_from_timed_sequence_diagram_with_symbolic_techniques_application_to_testing,termos_a_formal_language_for_scenarios_in_mobile_computing_systems,global_and_local_testing_from_message_sequence_charts} or Petri nets \cite{compositional_semantics_for_UML2_sequence_diagrams_using_petri_nets,automatic_model_transformation_from_uml_sequence_diagrams_to_coloured_petri_nets,a_toolset_for_conformance_testing_against_uml_sequence_diagrams_based_on_event_driven_colored_petri_nets}. 
Albeit those translations allow reusing advantageously the target framework's tools, relying on them to capture semantics leads to reasoning on foreign concepts. 
In \cite{timed_sequence_diagrams_and_tool_based_analysis_a_case_study}, UML-SDs are translated into timed automata, which are then verified with the UPPAAL tool \cite{uppaal_in_a_nutshell}. The translation mechanisms only concern models with synchronous communications. An observer automaton has to be designed so as to intercept communications between automata, make them observable, and enter an error state if other events are observed. 
In \cite{eliciting_unitary_constraints_from_timed_sequence_diagram_with_symbolic_techniques_application_to_testing}, each lifeline is translated into a timed input output symbolic transition system (TIOSTS) and message passing relies on some synchronous product. In order to cope with asynchronism, FIFO based communication schema have been introduced to ensure the consistency of executions on different lifelines. Also, dedicated variables have to be introduced to keep track of branching choices specified by $alt$ or $loop$ operators.
In \cite{termos_a_formal_language_for_scenarios_in_mobile_computing_systems}, a symbolic automaton is built from UML-SD specifications in the goal of analyzing traces by means of valid, invalid or inconclusive verdicts.
\cite{global_and_local_testing_from_message_sequence_charts} focuses on how to test Message Sequence Charts when the system is only partially observed. A translation into a network of asynchronous concurrent automata allows to define semantics through a product automaton as in \cite{eliciting_unitary_constraints_from_timed_sequence_diagram_with_symbolic_techniques_application_to_testing}.
In \cite{compositional_semantics_for_UML2_sequence_diagrams_using_petri_nets}, UML-SD specifications are translated into multivalued
nets (M-nets). The translation is compositional, entry and exit places of the M-nets corresponding to subinteractions being connected differently according to the parent combined fragment. However this process is complicated by the tracking of actions that are completely unordered w.r.t. one another. \cite{compositional_semantics_for_UML2_sequence_diagrams_using_petri_nets} also treats data in the form of variables, message parameters and guards. 
In \cite{automatic_model_transformation_from_uml_sequence_diagrams_to_coloured_petri_nets}, the authors propose an approach to automatically translate UML-SDs designed with the Papyrus tool \cite{papyrus_a_uml2_tool_for_domain_specific_language_modeling} to Coloured Petri Nets (CPNs) in a format compatible with CPNTools \cite{coloured_petri_nets_and_cpntools_for_modelling_and_validation_of_concurrent_systems}. 
CPNs come with an execution semantics that is particularly adapted for the description and analysis of distributed and concurrent systems. 
In \cite{automatic_model_transformation_from_uml_sequence_diagrams_to_coloured_petri_nets}, the translation revolves around a list of 11 rules with different priorities and which are applied to translate different concepts (lifelines, message occurrences, combined fragments, etc.) while iterating sequentially through the UML-SD's elements.
In \cite{a_toolset_for_conformance_testing_against_uml_sequence_diagrams_based_on_event_driven_colored_petri_nets} a set of UML-SDs are translated into Extended Petri Nets. Input execution traces can then be checked against the EPNs.

\noindent \textbf{Operational approach.} The literature contains few attempts at defining operational semantics for ILs. In \cite{operational_semantics_for_msc}, the authors build formal expressions over a process algebra signature. Starting from axioms such as $\epsilon \downarrow$ (the empty process $\epsilon$ terminates) and \resizebox{!}{10pt}{$a \xrightarrow{a} \epsilon$} ($a$ being an atomic action), an expression describing a MSC is build using rules such as \resizebox{!}{10pt}{$(x \xrightarrow{a} x')\wedge (y \not\xrightarrow{a}) \Rightarrow (x \mp y \xrightarrow{a} x')$}. Such an expression is then associated with a transition graph. The contribution in \cite{operational_semantics_for_msc} does not however deal with $loop$ operator and it is quite different from ours as the proposed transformations operate on process-algebraic expressions and not on syntactic terms. 
In contrast, the semantics proposed in \cite{a_fully_general_operational_semantics_for_UML2_sequence_diagrams_with_potential_and_mandatory_choice} relies on syntactic term transformations. Still, it also requires a communication medium as it is defined as the output of a combination of two transitions systems: an execution system which keeps track of communications, and a projection system which selects the next action to execute and provide the resulting interaction. As explained in \cite{a_hierarchy_of_communication_models_for_message_sequence_charts}, communication models keep track of emitted messages and messages pending receptions. They can for instance take the form of a set of dedicated buffers (e.g. FIFO). Our approach has the advantage of making such communication models implicit.

\noindent \textbf{Discussions.} 
Despite interaction languages specifying no synchronisation mechanisms between lifelines, several approaches that aim to implement tools, impose synchronisation points when entering and exiting combined operators and at decision points ($alt$, $opt$, $loop$) \cite{termos_a_formal_language_for_scenarios_in_mobile_computing_systems,eliciting_unitary_constraints_from_timed_sequence_diagram_with_symbolic_techniques_application_to_testing,compositional_semantics_for_UML2_sequence_diagrams_using_petri_nets,the_many_meanings_of_uml2_sd_a_survey} (although more recent works such as \cite{a_toolset_for_conformance_testing_against_uml_sequence_diagrams_based_on_event_driven_colored_petri_nets,a_fully_general_operational_semantics_for_UML2_sequence_diagrams_with_potential_and_mandatory_choice} do not). Although translation-based approaches have the benefit of allowing the use of the many existing analysis tools (UPPAAL \cite{uppaal_in_a_nutshell}, DIVERSITY \cite{an_end_to_end_framework_for_safe_software_development}, CPNTools \cite{coloured_petri_nets_and_cpntools_for_modelling_and_validation_of_concurrent_systems} etc.) we postulate that direct operational approaches such as ours facilitate features such as animation and debugging, becoming for the most part free-of-charge by-products of the analysis process.
\section{Conclusion\label{sec:conclusion}}

In this paper we proposed an operational semantics for ILs, aimed at trace validity analysis. This semantic is built upon a formal syntax for interaction terms and validated back-to-back w.r.t. a reference denotational semantics.
Our semantics is built on partial order relations induced on messages by the syntax. Those relations allow the identification of immediately executable actions. Pruning techniques then ensure a consistent semantics based on successive transformations of the form \resizebox{!}{10pt}{$i \xrightarrow{act} i'$}. 
On this principle, we have defined and implemented algorithms to compute semantics and to analyze the validity of traces. 
Experiments were successfully conducted in order to evaluate the correctness of each.

We intend to enrich our formalism: \textbf{(1)} by expanding trace analysis to a distributed context, where a set of traces (multi-trace) may be analyzed concurrently on a subset of observed lifelines; \textbf{(2)} by investigating whether or not our algorithmic treatments are fast enough to deal with traces on-the-fly so as to adapt them to monitoring. \textbf{(3)} by extending our IL to include modality operators such as $assert$ or $negate$. \textbf{(4)} by allowing the use of message arguments, variables, clocks and constraints within models.

Additionally, it would be interesting to perform a comparison with translation-based approaches. This may consist in a comparison of formal semantics and/or in benchmarking implementations according to a certain performance metric.

\clearpage

\bibliographystyle{splncs04}
\bibliography{revisiting_semantics_of_interactions_for_trace_validity_analysis}


\vfill

{\small\medskip\noindent{\bf Open Access} This chapter is licensed under the terms of the Creative Commons\break Attribution 4.0 International License (\url{http://creativecommons.org/licenses/by/4.0/}), which permits use, sharing, adaptation, distribution and reproduction in any medium or format, as long as you give appropriate credit to the original author(s) and the source, provide a link to the Creative Commons license and indicate if changes were made.}

{\small \spaceskip .28em plus .1em minus .1em The images or other third party material in this chapter are included in the chapter's Creative Commons license, unless indicated otherwise in a credit line to the material.~If material is not included in the chapter's Creative Commons license and your intended\break use is not permitted by statutory regulation or exceeds the permitted use, you will need to obtain permission directly from the copyright holder.}

\medskip\noindent\includegraphics{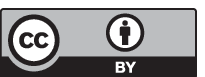}

\end{document}